**1A1**

14<sup>èmes</sup> Journées Nationales Microondes
11-12-13 Mai 2005 Nantes

# Amélioration de l'étude de l'humidité de sols par radiométrie.
# Caractérisation et modélisation diélectriques de profils géologiques.

F. Demontoux*, G.Ruffié*, J.P. Wigneron**, Maria-José Escorihuela***
* Laboratoire PIOM-ENSCPB-UMR 5501- 16 av Pey-Berland 33607 Pessac
**INRA-Unité de Bioclimatologie, BP 81, Villenave d'Ornon Cedex 33883
***CESBIO 18 avenue. Edouard Belin, bpi 2801 31401 Toulouse cedex 9

## I. Introduction

L'Humidité de surface du sol est une variable clé pour décrire les échanges d'eau et d'énergie entre la terre et l'atmosphère. En hydrologie et en météorologie, la quantité d'eau contenue dans les couches supérieures du sol (0-5cm de la surface) permet d'évaluer le rapport entre l'évaporation réelle et l'évaporation potentielle au niveau d'un sol nu. Il est aussi possible de déterminer la répartition des précipitations en eaux de ruissellements ou en eaux « stockées » ou d'autres variables comme la conductivité hydraulique.

Des études ont montré que des capteurs micro-ondes passifs pouvaient être utilisés pour scruter l'humidité de surface des sols. La solution qui a été retenue par l'équipe associée à la mission SMOS (Soil Moisture and Ocean Salinity) consiste à utiliser un radiomètre (1.4 GHz) [1] afin de relever les émissions micro-ondes des sols. Les mesures sont effectuées sur le site du CESBIO à Toulouse où un radiomètre à 1.4GHz est installé en permanence. Ces mesures permettent de connaître une permittivité équivalente des sols mesurés. Cependant, l'effet de la couverture de végétation, de la température du sol, de la couverture neigeuse, de la topographie ou des variations d'humidité joue un rôle considérable dans l'émission micro-ondes en surface. D'autres paramètres comme la présence d'inclusions dans le sol (trous, pierres) ou la texture du sol peuvent aussi perturber les mesures. Le but des travaux que nous présentons dans ce papier est de mettre au point un modèle numérique permettant de simuler des structures géologiques complexes. Ce modèle doit prendre en compte l'ensemble des paramètres pouvant influencer la permittivité équivalente mesurée par le radiomètre (état de surface, variation d'humidité, inclusions…). Le but de ce modèle est de calculer la permittivité équivalente de ces structures géologiques afin de pouvoir associer à chaque mesure du radiomètre une humidité équivalente.

## II. Caractérisation

La première partie de notre travail a consisté à effectuer des mesures en transmissions et en réflexions d'échantillons de sols agricoles placés dans un guide d'onde afin de connaître leur permittivité entre 1.3 GHz et 1.5 GHz.

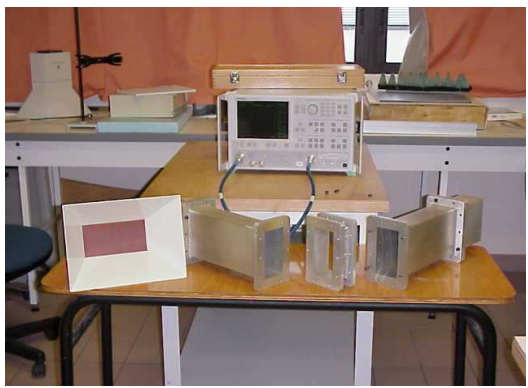

**Figure 1** : Banc de mesure en guide d'onde

La deuxième étape a consisté à mettre au point un procédé de caractérisation de sol. Dans un bac de $1m^3$ nous avons construit une structure géologique simple, que nous avons éclairée à l'aide d'un cornet. Une mesure en réflexion permet de remonter à la permittivité équivalente du milieu. Une fois le procédé validé, nous avons effectué des mesures in situ.

14<sup>èmes</sup> Journées Nationales Microondes, 11-12-13 Mai 2005 - Nantes

Ces résultats permettent de débuter l'étude de la variation de permittivité de ces différents milieux en fonction de l'humidité. Ils permettent aussi de valider les premiers résultats obtenus à l'aide de notre modèle.

### III. Développement du modèle

La finalité de ce modèle est de compléter les mesures afin de caractériser les structures géologiques [2]. Ce modèle doit permettre d'étudier la permittivité équivalente de sols en représentant la mesure expérimentale effectuée à l'aide du cornet. Il doit aussi prendre en compte de nombreux paramètres comme la topologie, la présence d'hétérogénéités (trous, pierres….) ou encore les variations d'humidité. Il doit aussi permettre d'étudier une surface suffisamment grande.

Les premières études ont été réalisées avec le logiciel HFSS-ANSOFT. Il repose sur la méthode des éléments finis. Il permet une première étude en prenant en compte un milieu homogène. Cette première étude simule le banc de mesure dont nous disposons au laboratoire. Il nous a permis d'étudier la réponse en réflexion de différents matériaux éclairés par un cornet (Figure 2).

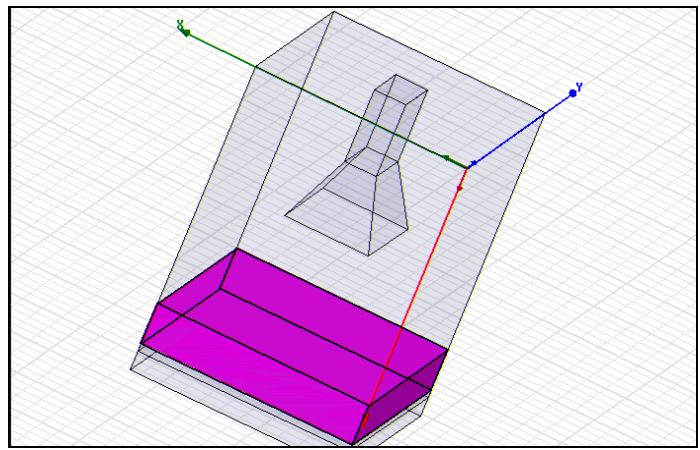

**Figure 2** : Banc de mesure en réflexion simulé

Les figures 3 présentent les résultats du module et de la phase du coefficient de réflexion normalisés par une mesure sur un court circuit. La comparaison des résultats expérimentaux et ceux issus de simulations montrent une bonne corrélation.

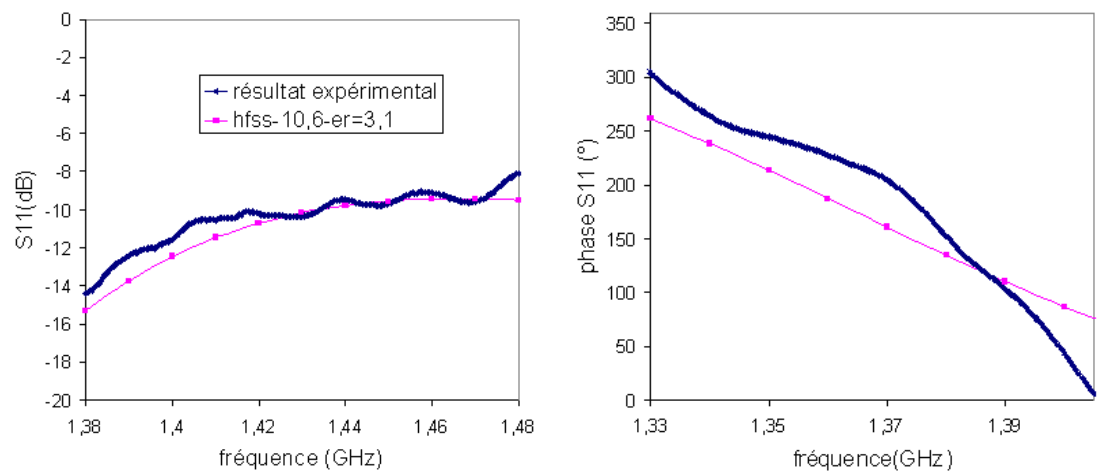

**Figure 3** : résultats de mesures et de calculs en réflexion

Des calculs complémentaires sur divers matériaux nous ont permis d'évaluer la sensibilité de la variation du paramètre S11 en fonction de la nature du produit.

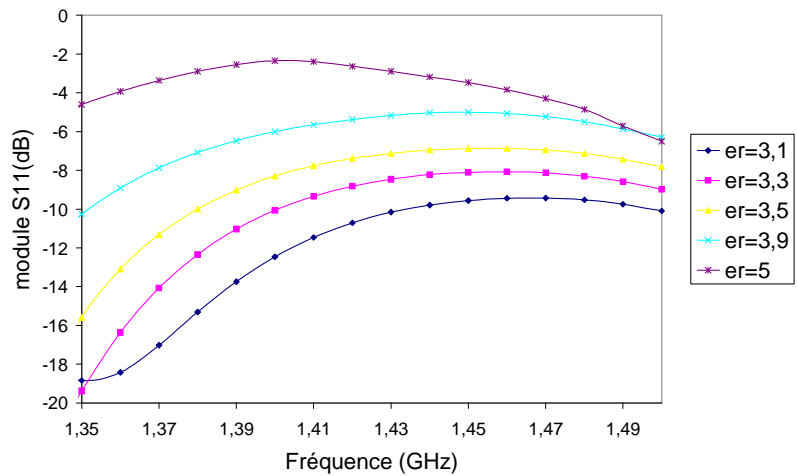

**Figure 4** : Simulations sur différents matériaux

Ces premières études nous donnent des informations très importantes sur les résultats que nous serons amené à traiter lorsque nous effectuerons des mesures sur des sols d'humidité très différentes. Malheureusement cette démarche ne nous permet pas encore d'évaluer l'influence de paramètres comme les gradients d'humidité, la présence d'air ou la nature non plane des couches étudiées. Pour introduire tous ces paramètres nous avons envisagé la mise au point d'un autre modèle de calcul.

La méthode de simulation que nous avons choisie pour ce nouveau modèle est la méthode FDTD [3]. C'est une méthode très connue qui a l'avantage de permettre l'introduction de l'ensemble des fonctions de notre modèle. Cette méthode a toutefois le défaut d'être très gourmande en mémoire à cause du grand nombre de mailles de calculs nécessaires pour simuler correctement notre problème. Ce nombre de maille important provient tout d'abord de la précision avec laquelle nous souhaitons représenter les formes des objets présents et d'un nombre minimum de cellules à utiliser par longueur d'onde pour une bonne application de l'algorithme de résolution. Ces problèmes peuvent être évités par l'utilisation d'un maillage à pas variable et d'un super calculateur SGI Altix/3300 à 12 processeurs et 16Go de RAM disponible à l'ENSCPB. Une étude d'une surface de 200 à 300 m$^2$ est alors possible.

Une autre limitation de cette méthode repose sur le type de source utilisée. Il est possible de représenter notre antenne cornet afin de simuler nos expériences. Malheureusement la prise en compte de l'antenne limite nos ressources mémoire pour représenter le reste de notre problème. Une autre approche consiste à générer un front d'onde plane sur notre structure géologique. Dans ce cas nous n'amputons pas nos ressources informatiques mais nous simplifions le comportement de notre antenne cornet et nous éloignons notre modèle de la réalité. Nous avons envisagé une troisième possibilité qui consiste à générer une source électromagnétique au niveau d'un plan source sans représenter l'antenne mais dont le comportement est voisin de celui du cornet (Figure 2). Le problème de maillage de l'antenne et d'utilisation d'une onde plane est alors contournée.

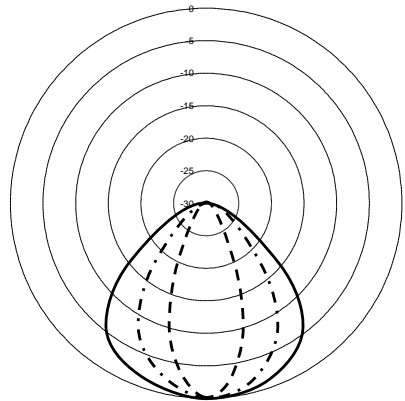

**Figure 5** : Variation du diagramme de rayonnement de notre source

### IV. Conclusion

Nous présenterons les résultats expérimentaux et ceux issus de notre modèle qui nous ont permis de débuter la caractérisation diélectrique de différentes structures géologiques en fonction de l'humidité.